\documentclass[11pt]{article}
\usepackage[pdftex]{color}
\usepackage{graphicx}
  \usepackage{amsthm,amsfonts}
  \usepackage{amsmath}
\usepackage{slashed}
\newcommand{\bea}   {\begin{eqnarray}}
\newcommand{\eea}   {\end{eqnarray}}

\begin{document}
\renewcommand{\thefootnote}{\fnsymbol{footnote}}

\thispagestyle{empty}

\title{On the spectrum-generating superalgebras of the \\deformed one-dimensional quantum oscillators}
\author{N. Aizawa\thanks{{E-mail: {\em aizawa@p.s.osakafu-u.ac.jp}}},\quad I. E. Cunha\thanks{{E-mail: {\em ivanec@cbpf.br}}}, \quad Z. Kuznetsova\thanks{{E-mail: {\em zhanna.kuznetsova@ufabc.edu.br}}}\quad and\quad
F.
Toppan\thanks{{E-mail: {\em toppan@cbpf.br}}}
\\
\\
}
\maketitle

\centerline{$^{\ast}$ {\it Department of Physical Sciences, Graduate School of Science,}}{\centerline{\it  Osaka Prefecture University, Nakamozu Campus, Sakai, Osaka 599-8531 Japan.}}
\centerline{$^{\dag\S}$
{\it CBPF, Rua Dr. Xavier Sigaud 150, Urca,}}\centerline{\it{
cep 22290-180, Rio de Janeiro (RJ), Brazil.}}
{\centerline{$^{\ddag}$ {\it UFABC, Av. dos Estados 5001, Bangu,}}}\centerline{\it { cep 09210-580, Santo Andr\'e (SP), Brazil.}}
~\\
\maketitle
\begin{abstract}
We investigate the dynamical symmetry superalgebras of the one-dimensional matrix superconformal quantum mechanics with inverse-square potential. They act as spectrum-generating superalgebras for the systems with the addition of the de Alfaro-Fubini-Furlan oscillator term.
The undeformed quantum oscillators are expressed by $2^n\times 2^n$ supermatrices; their corresponding spectrum-generating superalgebras are given by the $osp(2n|2)$ series. 
For $n=1$ the addition of a inverse-square potential does not break the $osp(2|2)$ spectrum-generating superalgebra. 
For $n=2$ two cases of inverse-square potential deformations arise. The first one produces Klein deformed quantum oscillators; the corresponding spectrum-generating superalgebras are given by the $D(2,1;\alpha)$ class, with $\alpha$ determining the inverse-square potential coupling constants.
The second $n=2$ case corresponds to deformed quantum oscillators of non-Klein type. In this case the
$osp(4|2)$ spectrum-generating superalgebra of the undeformed theory is broken to $osp(2|2)$.
The choice of the Hilbert spaces corresponding to the admissible range of the inverse-square potential coupling constants and
the possible direct sum of lowest weight representations of the spectrum-generating superalgebras is presented.  

~\\\end{abstract}
\vfill
\rightline{Revised version: March 8, 2019}
\rightline{CBPF-NF-004/18}

\newpage

\section{Introduction and summary}

In this paper we present a systematic investigation of the one-dimensional matrix oscillators deformed by diagonal
inverse-square potentials. We derive the general conditions for the existence of (one-dimensional, superconformal)
spectrum-generating superalgebras. We give the most general solutions (up to similarity transformations and with at least ${\cal N}=2$ supersymmetries) for $2\times 2$ and $4\times 4$ supermatrices. For these cases we compute the admissible Hilbert spaces and prove that,
depending on the range of the inverse-square potential coupling constants, the Hilbert space can be identified with a single
lowest weight representation of the spectrum-generating superalgebra or with a direct sum of its lowest weight
representations.  In the latter case the selection of the Hilbert space for the given model is not unique. This feature
was already noted in \cite{{mt},{ftf}} for the purely bosonic case of ordinary (i.e., not matrix) inverse-square potential quantum mechanics \cite{cal} and de  Alfaro-Fubini-Furlan oscillator \cite{dff}. In this work, among other results,  we extend the \cite{{mt},{ftf}} analysis to the (super)matrix case.\par
The one-dimensional $2^n\times 2^n$ undeformed matrix oscillators possess $osp(2n|2)$ spectrum generating superalgebras.  For $n=1$ the addition of a inverse-square potential does not break the $osp(2|2)$ spectrum-generating superalgebra. 
For $n=2$ two cases of inverse-square potential deformations arise. The first one produces Klein deformed quantum oscillators \cite{vas} (see also \cite{ply} and references therein for Klein oscillators); the corresponding spectrum-generating superalgebras are given by the $D(2,1;\alpha)$ class, with $\alpha$ determining the inverse-square potential coupling constants.
The second $n=2$ case corresponds to deformed quantum oscillators of non-Klein type. In this case the
$osp(4|2)$ spectrum-generating superalgebra of the undeformed theory is broken to $osp(2|2)$.\par
The topic under investigation is receiving considerable attention in the literature. Three frameworks are presently used for
investigations. The most popular one consists in the quantization of (superconformal) worldline sigma-models; a second approach consists in analyzing symmetries (following, e.g., \cite{olv}) of partial differential equations which,
for the present case, are time-dependent Schr\"odinger equations of matrix type; the third approach is the
one here employed. The main motivations for investigating classical worldline superconformal sigma-models (and their
quantization) come from the recognition that they underlie the dynamics of test particles in the proximity of the horizon of certain black holes (see \cite{bp}) and for their role in the $AdS_2/CFT_1$ correspondence \cite{{sen},{cj}}.
The worldline superconformal sigma-models (for a review see \cite{fil} and references therein)  are obtained by imposing constraints to the supersymmetric sigma-models associated with one-dimensional supermultiplets \cite{{pato},{kuroto}}. They can be derived by using superspace \cite{fil} or $D$-module representations of superconformal algebras \cite{{kuto},{khto}}. In \cite{hoto}, extending the construction pointed out in \cite{pap}, it was shown  that
superconformal dynamical symmetries of worldline sigma-models are obtained from either parabolic or
trigonometric $D$-module representations of the superconformal algebra. The first case corresponds to the classical version of the inverse-square potential, while the second case corresponds to the classical addition of the de Alfaro-Fubini-Furlan oscillator term. The quantization of the parabolic superconformal sigma-models has been performed in several
papers (see, e.g., \cite{{fil2},{fil3}} for the $D(2,1;\alpha)$ superconformal models). The quantization of trigonometric superconformal sigma-models has been done
in fewer works (in \cite{{ivsi},{ils}} for undeformed oscillators, while the first example of deformed oscillator was produced in
\cite{cuhoto}). The quantization of these systems is given in terms of quantum Noether charges which are expressed in the Heisenberg framework. We point out that the above papers lack the full analysis of the selection of the admissible Hilbert spaces as done in \cite{{mt},{ftf}} and the present work.\par
The approach based on the symmetry of a matrix partial differential equation was discussed in \cite{tv} for a specific model. The deformed oscillator system described in Section {\bf 4} is the time-independent version of the time-dependent Schr\"odinger equation introduced in \cite{tv}.  The anaysis of this model is made here more explicit in three points, namely  the recognition  that the $osp(1|2)\subset osp(2|2)$ subalgebra is sufficient to determine the spectrum of the theory,  the selection of the admissible Hilbert spaces in the three different intervals
($\beta\leq -\frac{1}{2}$, $-\frac{1}{2}<\beta<\frac{1}{2}$, $\beta\geq \frac{1}{2}$) of the deformation parameter $\beta$ and, finally, the computation of the orthonormal eigenstates.\par
The model with $D(2,1;\alpha)$ spectrum-generating superalgebra discussed in Section {\bf 5} was first derived in
\cite{cuhoto}. We present here a more thorough analysis which includes the selection of the admissible Hilbert spaces for the three different intervals of the deformation parameter $\alpha$, as well as the recognition that the $osp(2|2)\subset
D(2,1;\alpha)$ subalgebra is sufficient to determine the spectrum of the theory. This extra analysis is made possible by the simplification which occurs in presenting the operators of the spectrum-generating superalgebra in 
a Schr\"odinger framework (therefore, with no time dependence) instead of the Heisenberg framework of  \cite{cuhoto}.  
\par
The deformed oscillator (whose spectrum and orthonormal eigenstates have been computed) introduced in Section {\bf 6}  is a genuine new model. It is the simplest example of a inverse-square deformed oscillator of non-Klein type.
Its spectrum-generating superalgebra is $osp(2|2)$. The model depends on a real deformation parameter $\nu$. The Hilbert space exists (and is unique, being given by a lowest weight representation of $osp(2|2)$) for $\nu\neq 0$. The  vacuum is unique,
while all excited states are doubly degenerate, so that the semiinfinite $(1,2,2,2,\ldots )$ tower of states is produced.
Unlike the Klein-deformed matrix oscillators (\ref{hoscn1},\ref{osckleinn2}), the non-Klein deformed matrix oscillator (\ref{nklosc}) is not obtained as a continuous deformation of the undeformed oscillator. \par
The scheme of the paper is as follows. In Section {\bf 2} we derive the spectrum-generating superalgebras of the
undeformed one-dimensional  matrix oscillators. In Section {\bf 3} we discuss their deformations via the introduction of diagonal inverse-square potentials.  The $n=1$ example of deformed oscillators is presented in Section {\bf 4}.  In Section {\bf 5} the analysis is extended to the Klein deformations of the $n=2$ oscillators. In Section {\bf 6} we produce the results for the non-Klein deformation
of the $n=2$ matrix oscillators.  Some comments about the present knowledge of the existing deformations for $n\geq 3$ are given in Section {\bf 7}. In the Conclusions we mention open problems for further investigations.  The paper is complemented by three Appendices. In Appendix {\bf A} we discuss
 the relevant features of the subclass of finite Lie superalgebras which are one-dimensional superconformal. In Appendix {\bf B} we present the basic properties of the exceptional class of $D(2,1;\alpha)$ superalgebras. Finally, in Appendix {\bf C},
we discuss the selection of the admissible Hilbert spaces  for the de Alfaro-Fubini-Furlan deformed oscillators. 
 
\section{The undeformed one-dimensional quantum oscillators and their $osp(2n|2)$ spectrum-generating superalgebras}

As recalled in \cite{akt}, the  (\ref{sqm}) superalgebra (see Appendix {\bf A}) of the supersymmetric quantum mechanics can be  constructed by Hermitian matrix differential operators $Q_I$, $H$ acting on a supermultiplet of real-valued fields. On the other hand
the introduction of a dynamical symmetry realized by Hermitian operators closing a superconformal algebra requires
a complex structure. The reason is the presence of non-vanishing commutators (such as
$[Q_I,K]=i{\widetilde Q}_I$); they imply that the imaginary unit has to be introduced in order to have Hermitian operators on the right hand side. Therefore, without loss of generality,  we can investigate superconformal dynamical symmetries (and spectrum-generating superalgebras) acting on supermultiplets of complex fields.\par
The one-dimensional $2^n\times 2^n$ free matrix  Hamiltonian $H$ is given by
\bea\label{freeham}
H&=&-\frac{1}{2}\partial_x^2 \cdot{\mathbb I}_{2^{n}}
\eea
(here and in the following ${\mathbb I}_k$ denotes the $k\times k$ identity matrix). 
\par
For any positive integer $n\in {\mathbb N}$, $H$ possesses $2n$ distinct Hermitian, fermionic (i.e. block-antidiagonal) first-order matrix differential operators $Q_I$ as its square roots. The $Q_I$ operators close the ${\cal N}$-extended superalgebra (\ref{sqm}) with
\bea
{\cal N} &=& 2n.
\eea
The above relation between ${\cal N} $ and $2n$ is based on the constructions reported in \cite{oku,crt} for complex-valued Clifford algebras.  There are $2n$ block antidiagonal complex matrices
$\gamma_I$, $I=1,2,\ldots, 2n$, satisfying the relations 
\bea
\gamma_I\gamma_J+\gamma_J\gamma_I = 2\delta_{IJ} \cdot{\mathbb I}_{2^{n}}.
\eea
 The extra block-diagonal matrix $F$,
\bea\label{Fparity}
F&=& {\small{\left(\begin{array}{cc}{\mathbb I}_{2^{n-1}}&0\\0&-{\mathbb I}_{2^{n-1}}\end{array}\right)}},
\eea
satisfies the anticommutation relations
\bea
F\gamma_I +\gamma_I F&=&0, \quad\quad \forall I=1,2,\ldots, 2n.
\eea
$F$ is called the fermion parity operator. Its eigenvectors with $+1$ ($-1$) eigenvalue are the even, also called bosonic (odd, also called fermionic),  states.\par
We can set
\bea
Q_I &=& \frac{i}{\sqrt{2}}\gamma_I \partial_x,
\eea
so that (\ref{sqm}) reads as
\bea\label{sqmfree}
&\{Q_I, Q_J\}= 2\delta_{IJ} H, \quad\quad [H,Q_I] =0,\quad\quad  {\text{for}}\quad I,J=1,\ldots, {2n}.&
\eea
\par
The conformal counterpart of the Hamiltonian $H$ is the oscillator $K$ which can be assumed to be proportional to the identity matrix. Therefore
\bea\label{oscop}
K&=&\frac{1}{2} x^2 \cdot {\mathbb I}_{2^n}.
\eea 
The conformal counterparts of the $Q_I$ operators are the Hermitian operators ${\widetilde Q}_I$, introduced
through
\bea
[Q_I, K] = i {\widetilde Q}_I   &\rightarrow & {\widetilde Q}_I = \frac{1}{\sqrt{2}} x\cdot \gamma_I.
\eea
The dilatation operator $D$ and the $R$-symmetry operators $\Sigma_{IJ}=-\Sigma_{JI}$ are introduced
from the anticommutators
\bea
\{ Q_I, {\widetilde Q}_J\} &=& -2\delta_{IJ} D + \Sigma_{IJ}.
\eea
We have
\bea
D&=& -\frac{i}{2}(x\partial_x+\frac{1}{2})\cdot {\mathbb I}_{2^n}, \nonumber\\
\Sigma_{IJ} &=& ~~\frac{i}{2}\gamma_I\gamma_J.
\eea
For any positive integer $n$ the set of Hermitian operators $D,H, K, Q_I, {\widetilde Q}_I,\Sigma_{IJ}$ close the $D(n,1)\sim osp(2n|2)$ superalgebra. The $4n$ generators $Q_I$, ${\widetilde Q}_I$ are odd. The $n(2n-1)+3$ even generators
$H,D,K, \Sigma_{IJ}$ produce the $sl(2)\oplus so(2n)$ subalgebra. The superalgebra $osp(2n|2)$ belongs to the class of one-dimensional superconformal algebras discussed in Appendix {\bf A} (the $so(2n)$ subalgebra is the $R$-symmetry). \par
We present, for completeness, the non-vanishing  (anti)commutators of $osp(2n|2)$. In order to write them in more compact form we introduce the generators $E^+ = H$, $E^-=K$, $Q_I^+=Q_I$, $Q_I^-={\widetilde Q}_I$. We have
\bea
\relax [D, E^\pm] =\pm i E^\pm,~&& ~~~[D,Q_I^\pm]= \pm\frac{i}{2}Q_I^\pm,\nonumber\\
\relax [E^+,E^-] =-2iD,~&& \{Q_I^+, Q_I^-\}= -2\delta_{IJ} D +\Sigma_{IJ},\nonumber\\
\relax [ E^\pm, Q_I^\mp]=\pm i Q_I^\pm,~&& \{Q_I^\pm, Q_J^\pm\}=2\delta_{IJ}E^\pm,\nonumber \\
\relax~~ [\Sigma_{IJ},\Sigma_{IL} ]= -i\Sigma_{JL},&&  ~[\Sigma_{IJ}, Q_K^\pm]=-i\delta_{IK} Q_J^\pm+i\delta_{JK}Q_I^\pm .
\eea
The Hamiltonian $H_{osc}$  of the (undeformed) matrix oscillator is the sum of $H$ and $K$:
\bea\label{hosc}
H_{osc}&=& H+K = \frac{1}{2}(-\partial_x^2+x^2)\cdot {\mathbb I}_{2^n}.
\eea
The superalgebra $osp(2n|2)$ is the spectrum-generating superalgebra for $H_{osc}$. A linear combination of
the odd generators produce $2n$ pairs of $a_I^\dagger$, $a_I$ creation/annihilation operators satisfying $2n$ independent Heisenberg algebras defined by their commutators.\par
We can set
\bea
a_I=Q_I+i{\widetilde Q}_I=\frac{i}{\sqrt 2}\gamma_I(\partial_x+x),&& a^\dagger_I= Q_I-i{\widetilde Q}_I=\frac{i}{\sqrt 2}\gamma_I(\partial_x-x).
\eea
In terms of anticommutators we have (no summation over the repeated indices is understood)
\bea
H_{osc} &=& \frac{1}{2}\{a_I,a_I^\dagger\}.
\eea
$a_I^\dagger$ ($a_I$) are creation (annihilation) operators due to the commutators
\bea
\relax [H_{osc}, a_I^\dagger]=a_I^\dagger,&\quad&[H_{osc}, a_I]=-a_I.
\eea
For every $I$, the Heisenberg algebras are recovered from
\bea
\relax [a_I,a_I^\dagger]&=& {\mathbb I}_{2^n}.
\eea
The annihilation operators $a_I$ allow to define the $2^n$ degenerate ground states $|0\rangle_I$ of $H_{osc}$ 
as the lowest weight vectors satisfying, for each $I$,
\bea
a_I|0\rangle_I=0, &\quad& H_{osc}|0\rangle_I= \frac{1}{2}|0\rangle_I.
\eea
Half of the degenerate ground states are bosonic and half of them are fermionic. The Hilbert space of the undeformed matrix oscillator is a $2^n$-ple of ${\cal L}^2({\mathbb R})$ square integrable functions on the line.\par

\section{The inverse-square potential in matrix quantum Hamiltonians}

The addition to the free Hamiltonian $H$ in (\ref{freeham})  of a inverse-square potential $\frac{1}{x^2}V$, where 
$V=diag(v_1,v_2,\ldots v_{2^n})$ is a $2^n\times 2^n$ constant diagonal matrix,  is such to preserve the scaling property of
$H$. Indeed, if we set 
the scaling dimension of the space coordinate to be $[x]=-1$, then 
\bea
&[H] =[H+\frac{1}{x^2}V] = 2.&
\eea
In this paper we address the question of the constraints to be imposed on the inverse-square potential coupling constants $v_i$'s entering the
diagonal matrix $V$ in order to get a one-dimensional superconformal Lie algebra as a dynamical symmetry of
the inverse-square deformed Hamiltonian $H_{def}$, defined as
\bea
H_{def}&=&H+\frac{1}{x^2}V.
\eea 
By construction the associated inverse-square deformed oscillator $H_{osc}+\frac{1}{x^2}V$, with $H_{osc}$ given in (\ref{hosc}),  possesses the obtained one-dimensional superconformal Lie algebra as a spectrum-generating superalgebra.\par
Obviously the $osp(2n|2)$ dynamical symmetry
of the free Hamiltonian is  in general no longer a dynamical symmetry of the $\frac{1}{x^2}V$ inverse-square deformed Hamiltonian. It is worth noticing, on the other hand, that the dynamical symmetry of the inverse-square deformed Hamiltonian is not necessarily a subalgebra of $osp(2n|2)$, as one could na\"ively expect. In some cases (discussed in the following in Sections {\bf 5} and {\bf 7}) it corresponds to a deformation of $osp(2n|2)$.\par
The deformed supersymmetry operators have to be expressed as
 \bea\label{defsusyop}
Q_I ^{def}&=& \frac{i}{\sqrt{2}}(\gamma_I \partial_x-i\frac{M_I}{x}) ,
\eea
where the $M_I$'s should be block-antidiagonal, constant matrices satisfying the hermiticity condition $M_I^\dagger=M_I$.
\par
The closure of the (\ref{sqm}) superalgebra requires the following equations to be satisfied for $I\neq J$
\bea\label{Mrelations}
\{\gamma_I,M_J\}+\{\gamma_J, M_I\}&=&0,\nonumber\\
\{M_I,M_J\}-i\gamma_IM_J-i\gamma_JM_I&=&0.
\eea
At $I=J$ the potential of the inverse-square potential deformed Hamiltonian $H_{def}$ should be given by $\frac{1}{x^2}V$ where, for any $I$, we get
\bea\label{Vresult}
V&=& \frac{1}{2}(M_I^2-i\gamma_IM_I).
\eea
The (\ref{oscop}) oscillator operator $K$  remains undeformed; it follows that the dilatation operator $D$ and the fermionic operators ${\widetilde Q}_I$ are also unchanged.\par
In order to recover the dilatation operator $D$ from the anticommutator $\{Q_I^{def},{\widetilde Q}_I\}$, for any $I$ the condition
\bea\label{Dconstraint}
\{M_I,\gamma_I\}&=&0
\eea 
has to be fulfilled.\par
The anticommutators $\{Q_I^{def},{\widetilde Q}_J\}$ for $I\neq J$ give the constant operators
\bea\label{sigmadef}
\relax \Sigma_{IJ}^{def} &=& \frac{1}{2}(\frac{i}{2}[\gamma_I,\gamma_J] +\{M_I,\gamma_J\}).
\eea
Since the first relation in (\ref{Mrelations}) is assumed to be satisfied, then $\Sigma_{IJ}^{def}=-\Sigma_{JI}^{def}$.\par
The closure of a superconformal algebra is obtained provided that the $\Sigma_{IJ}^{def}$'s close the $R$-symmetry
subalgebra and that the fermionic operators $Q_I^{def}$, ${\widetilde Q}_I$ belong to $R$-symmetry representation multiplets.\par
A class of solutions of the (\ref{Mrelations},\ref{Vresult},\ref{Dconstraint}) constraints  is obtained by setting
\bea\label{kleindefsol}
M_I&=& i\beta \gamma_I F,
\eea
where $\beta$ is an arbitrary real number and $F$ is the fermion parity operator introduced in (\ref{Fparity}).
The (\ref{kleindefsol}) solution fails, however, to produce a superconformal algebra for $n\geq 3$.\par
The $\Sigma_{IJ}^{def}$ operators from (\ref{sigmadef}), under (\ref{kleindefsol}) deformation, read
\bea
\Sigma_{IJ}^{def}&=& \frac{i}{2}\gamma_I\gamma_J(1-2\beta F).
\eea
It is immediate to check that, for $n=1$, the introduction of the (\ref{kleindefsol}) deformation does not spoil
the $osp(2|2)$ dynamical symmetry of the free system.\par
For $n=2$ the closure of a superconformal algebra as a dynamical symmetry is guaranteed by the fact that $F$ is expressed by the product $F=\gamma_1\gamma_2\gamma_3\gamma_4$. This relation implies that the commutators
$[\Sigma_{IJ}^{def},{\widetilde Q}_K]$ close on the ${\widetilde Q}_L$ generators on the right hand side. A similar property
holds for the $Q_K$ generators.\par
The models based on the (\ref{kleindefsol}) deformation for $n=1$ and $n=2$ are explicity discussed in Section {\bf
4} and, respectively, Section {\bf 5}. \par
The (\ref{kleindefsol}) deformation implies that the deformed creation and annihilation operators satisfy a Klein-deformed Heisenberg algebra. We recall (see \cite{ply}) that the Klein-deformed Heisenberg algebra is realized by
a pair of Hermitian conjugated operators $a_{Kl}, a_{Kl}^\dagger$ satisfying the relations
\bea\label{kleinalg}
[a_{Kl}, a_{Kl}^\dagger]&=& \mathbb{I} + \nu {\overline K},\nonumber\\
\{a_{Kl},{\overline K} \}&=& \{a_{Kl}^\dagger,{\overline K}\} ~~=~~0,\nonumber\\
{\overline K}^2 &=& {\mathbb I},
\eea
for some given real number $\nu$. The operator ${\overline K}$, which is a square root of the Identity, is known as
``Klein operator".\par
In term of the (\ref{kleindefsol}) deformation we can set
\bea\label{kleinosc}
a_{Kl,I}&=& Q_I^{def}+i{\widetilde Q}_I= \frac{i}{\sqrt{2}}\gamma_I(\partial_x+x+\frac{\beta F}{x}),\nonumber\\
a_{Kl,I}^\dagger&=& Q_I^{def}-i{\widetilde Q}_I= \frac{i}{\sqrt{2}}\gamma_I(\partial_x-x+\frac{\beta F}{x}).
\eea
It follows 
\bea
[a_{Kl,I},a_{Kl,I}^\dagger]&=& {\mathbb I}-2\beta F,  \quad\quad \{a_{Kl,I},F\}=\{a_{Kl,I}^\dagger, F\}=0.
\eea
For any $I$ the deformed creation/annihilation operators  (\ref{kleinosc}) define a (\ref{kleinalg}) Klein-deformed Heisenberg algebra with $\nu=-2\beta$ and fermion parity operator $F$ as Klein operator.\par
Furthermore, we get the deformed oscillator Hamiltonian $H_{osc}^{Kl}$ from the anticommutators
\bea
\frac{1}{2} \{a_{Kl,I},a_{Kl,I}^\dagger\} &=& H_{osc}^{Kl}= H_{def}+K =\frac{1}{2}(-\partial_x^2 +x^2+ \frac{\beta^2+\beta F}{x^2}){\mathbb I}.
\eea
The Klein-deformed oscillators are creation/annihilation operators since
\bea
[H_{osc}^{Kl},a_{Kl,I}] =-a_{KL,I}, &&
[H_{osc}^{Kl},a_{Kl,I}^\dagger] =a_{KL,I}^\dagger.
\eea 
We close this Section by pointing out that the constraints (\ref{Mrelations},\ref{Vresult},\ref{Dconstraint}) admit more general solutions, different from the ones given by (\ref{kleindefsol}).  These solutions can also induce superconformal algebras as dynamical symmetries. One of such examples, leading to deformed creation/annihilation oscillators which do not satisfy the Klein condition, is presented in Section {\bf 6}.

\section{The $n=1$ case with Klein deformed oscillators and $osp(2|2)$ spectrum-generating superalgebra}

In this Section we present the $n=1$ Klein deformed oscillator. We show that its spectrum-generating superalgebra is $osp(2|2)$, like the undeformed case. The construction of the admissible Hilbert spaces is given at the end.\par
For $n=1$  the formulas of the operators given in Section {\bf 3} are specialized in terms of the three Pauli matrices 
$\sigma_i$, $i=1,2,3$, given by
\bea\label{pauli}
&\sigma_1=\tiny{\left(\begin{array}{cc}0&1\\1&0\end{array}\right)}, \quad\quad\quad
\sigma_2=\tiny{\left(\begin{array}{cc}0&-i\\i&0\end{array}\right)},\quad\quad\quad
\sigma_3=\tiny{\left(\begin{array}{cc}1&0\\0&-1\end{array}\right)}.&
\eea
For any real value of the parameter $\beta$ the four even operators $H,D,K,J$ and the four odd operators $Q_1,Q_2,{\widetilde Q}_1, {\widetilde Q}_2$ close the $osp(2|2)$ superalgebra. Their respective expressions are
\bea\label{osp22op}
H&=&\frac{1}{2}(-\partial_x^2 +\frac{\beta^2+\beta \sigma_3}{x^2})\cdot{\mathbb I}_2,\nonumber\\ 
D&=& -\frac{i}{2}(x\partial_x+\frac{1}{2})\cdot {\mathbb I}_{2}, \nonumber\\
K&=&\frac{1}{2}x^2\cdot{\mathbb I}_2,\nonumber\\
J &=& -\frac{1}{2}\sigma_3+\beta{\mathbb {I}}_2,\nonumber\\
Q_I&=& \frac{i}{\sqrt{2}}\sigma_I\cdot(\partial_x+\frac{\beta\sigma_3}{x}),\nonumber\\
{\widetilde Q}_I &=& \frac{1}{\sqrt{2}}\sigma_I\cdot x,
\eea
where $I=1,2$.\par
Their non-vanishing (anti)commutators are given in Appendix {\bf A}, formula (\ref{osp22structureconstants}).\par
One should note that the closure of the $osp(2|2)$ superalgebra is not affected by the presence of the non-vanishing
real parameter $\beta$.  The reality condition on $\beta$ is imposed to guarantee the hermiticity property of the (\ref{osp22op}) operators.\par
The Klein-deformed oscillators are introduced through the positions
\bea\label{n1klein}
a_{I}= Q_I+i{\widetilde Q}_I, && 
a_{I}^\dagger= Q_I-i{\widetilde Q}_I.
\eea
Therefore we obtain, for $I=1,2$,
\bea
a_I&=&\frac{i}{\sqrt 2}\sigma_I\cdot(\partial_x+\frac{\beta\sigma_3}{x}+x),\nonumber\\
a_I^\dagger &=& \frac{i}{\sqrt 2} \sigma_I\cdot (\partial_x +\frac{\beta \sigma_3}{x}-x).
\eea
At a given $I=1,2$, the commutator is
\bea\label{kleincomm}
\relax [a_I, a_I^\dagger] &=& {\mathbb I}_2-2\beta \sigma_3,
\eea
while the Hamiltonian $H_{osc}$ of the deformed oscillator is 
\bea\label{hoscn1}
&H_{osc}=\frac{1}{2}\{a_I,a_I^\dagger\}= H+K= \frac{1}{2}(-\partial_x^2+ x^2+ \frac{\beta^2+\beta \sigma_3}{x^2})\cdot{\mathbb I}_2.&
\eea
The condition
\bea
a_I|\lambda \rangle &=& 0
\eea
defines a lowest weight vector. For any real $\beta$ there are two such lowest weight vectors, one bosonic
($|\lambda_{Bos}\rangle$, such that $\sigma_3|\lambda_{Bos}\rangle=|\lambda_{Bos}\rangle$) and one fermionic
($|\lambda_{Fer}\rangle$, such that $\sigma_3|\lambda_{Fer}\rangle=-|\lambda_{Fer}\rangle$).\par
We have that
\bea
|\lambda_{Bos}\rangle \propto \left(\begin{array}{c} x^{-\beta} e^{-\frac{1}{2}x^2}\\0\end{array}\right), &&
|\lambda_{Fer}\rangle \propto \left(\begin{array}{c} 0\\ x^\beta e^{-\frac{1}{2}x^2}\end{array}\right).
\eea
The annihilation operator $a_2$ defines, up to a phase, the same lowest weight vectors as $a_1$. This property remains true for the excited states: $(a_2^\dagger)^n|\lambda_{Bos}\rangle$ ($(a_2^\dagger)^n|\lambda_{Fer}\rangle$) differs from $(a_1^\dagger)^n|\lambda_{Bos}\rangle$ ($(a_1^\dagger)^n|\lambda_{Fer}\rangle$) by a phase. It turns out that, as a spectrum-generating superalgebra of the $H_{osc}$ (\ref{hoscn1}) Hamiltonian, $osp(2|2)$ is redundant. The spectrum of the theory can be recovered from each one of the two copies of the $osp(1|2)\subset osp(2|2)$ subalgebras, either the one given by the generators $H,D,K, Q_1, {\widetilde Q}_1$, or the one given by the generators $H,D,K, Q_2, {\widetilde Q}_2$.\par
This important point deserves to be duly emphasized. We therefore present the\\
{\bf Remark:} {\em  the spectrum-generating superalgebra $osp(2|2)$ is redundant to produce the spectrum of the theory 
since the ray vectors of a Hilbert space are determined by the $osp(1|2)$ spectrum-generating subalgebra. }
\par
Repeating the analysis discussed in Appendix {\bf C} to the present case, we easily conclude that the lowest weight representation induced by $|\lambda_{Bos}\rangle$ defines a normed Hilbert space given by a pair of ${\cal L}^2({\mathbb R})$ square integrable functions on the real line, provided that the normalization condition 
$-2\beta>-1$ is satisfied. Similarly, $|\lambda_{Fer}\rangle$ defines a Hilbert space provided that the condition
$2\beta>-1$ is satisfied.\par
We arrive at the following selection of admissible Hilbert spaces for the model:
\begin{description}
\item[i)] in the range  $\beta \leq-\frac{1}{2}$ the only admissible Hilbert space corresponds to a single lowest weight representation, with $|\lambda_{Bos}\rangle$ as ground state. Its vacuum energy $E_{Bos}(\beta)$ is 
given by $E_{Bos}(\beta) =\frac{1}{2}-\beta$;
\item[ii)] in the range  $\beta \geq\frac{1}{2}$ the only admissible Hilbert space corresponds to a single lowest weight representation, with $|\lambda_{Fer}\rangle$ as ground state. Its vacuum energy $E_{Fer}(\beta)$ is 
given by $E_{Fer}(\beta)=\frac{1}{2}+\beta$;
\item[iii)] in the intermediate range $-\frac{1}{2}<\beta<\frac{1}{2}$ different choices of Hilbert space are admissible.
Either
\item{iiia)}  one can select as Hilbert space a single lowest weight representation (the lowest weight vector being either $|\lambda_{Bos}\rangle$ or $|\lambda_{Fer}\rangle$). Alternatively,
\item{iiib)}  the Hilbert space can be selected to be the direct sum of the two
lowest weight representations. The energy difference $\Delta(\beta)=E_{Bos}(\beta)-E_{Fer}(\beta)$ of the two ground states is $\Delta(\beta)= -2\beta$. Therefore, $|\lambda_{Bos}>$ is the vacuum state for $0<\beta<\frac{1}{2}$,
while $|\lambda_{Fer}\rangle$ is the vacuum state for $-\frac{1}{2}<\beta<0$. A degenerate ground state is recovered for the $\beta=0$ undeformed oscillator.
 \end{description}
We conclude this Section by pointing out that, without loss of generality, one can restrict the real parameter $\beta$ to belong to a half line (either $\beta\geq 0$ or $\beta\leq 0$). The reason for that is the existence of a similarity transformation, induced by the Pauli matrix $\sigma_1$, which allows to exchange bosonic and fermionic states. Under this similarity transformation any operator $g$ entering (\ref{osp22op}) is mapped into
\bea\label{simtra}
g'&=& \sigma_1g\sigma_1.
\eea
Let us stress the $\beta$-dependence of $H$ entering (\ref{osp22op}) by denoting it as ``$H(\beta)$". We obtain,
in particular, that the following relation is satisfied
\bea
 &H'(\beta)=\sigma_1H(\beta)\sigma_1= H(-\beta).&
\eea
In the range $0<\beta<\frac{1}{2}$ the vacuum state is $|\lambda_{Bos}\rangle$. In terms of the iiib) option for
the Hilbert space, the spectrum is given by
\bea
E_{\epsilon,n}&=& \frac{1}{2} -\epsilon \beta +n,
\eea
where $n\in{\mathbb N}_0$ and $\epsilon = \pm 1$. The vacuum energy corresponds to $\epsilon=1$, $n=0$.\par
Each energy level $E_{\epsilon,n}$ is not degenerate. The parity $P_{\epsilon,n}$  (even or odd) of the corresponding eigenfunctions,
given by the $\pm 1$ eigenvalues of the fermion parity operator $\sigma_3$, is given by
\bea
P_{\epsilon,n}&=&\epsilon(-1)^{n}.
\eea
We compute now the orthonormality conditions for the corresponding eigenfunctions in the range $0<\beta<\frac{1}{2}$ (the orthonormality conditions for the Klein-deformed operators were presented in \cite{ply} and references therein). Let us denote
as $|0\rangle_{\epsilon}$ the bosonic ($\epsilon=1$) and the fermionic ($\epsilon=-1$) normalized lowest weight states, so that $~_\epsilon\langle 0|0\rangle_\epsilon=1$. We determine $N_{\pm 1}$ so that
\bea
|0\rangle_{1}= N_1 \left(\begin{array}{c} x^{-\beta} e^{-\frac{1}{2}x^2}\\0\end{array}\right), &\quad&
|0\rangle_{-1} = N_{-1} \left(\begin{array}{c} 0\\ x^\beta e^{-\frac{1}{2}x^2}\end{array}\right).
\eea
They are determined by the conditions, see \cite{akt},
\bea
|N_\epsilon|^2\int_{-\infty}^{+\infty} dx| x^{-2\epsilon\beta}e^{-x^2}| &=&1.
\eea
We express the line integral in terms of the Gamma function. At first we separate the line integral into two integrals: $\int_{-\infty}^{+\infty} =\int_{-\infty}^0+\int_0^{+\infty}$. By changing the integration variable ($x\mapsto -x$) in the first integral on the right hand side we are able to write $\int_{-\infty}^{+\infty}dx| x^{-2\epsilon \beta}e^{-x^2}| = (1+|(-1)^{-2\epsilon\beta}|)\int_0^{+\infty} dx x^{-2\epsilon \beta} e^{-x^2}= 2\int_0^{+\infty}dx x^{-2\epsilon\beta}e^{-x^2}$. With the further change of the integration variable by setting $t=x^2$ we obtain $2\int_0^{+\infty}dx x^{-2\epsilon\beta}e^{-x^2}= \int_0^{+\infty}dt t^{-\epsilon\beta-\frac{1}{2}}e^{-t}= \Gamma(-\epsilon\beta+\frac{1}{2})$.
The normalization factors $N_\epsilon$ can therefore be expressed as
\bea\label{npm1}
N_\epsilon &=& \frac{1}{\sqrt{\Gamma(-\epsilon\beta+\frac{1}{2})}}.
\eea 
The unnormalized excited states $|{\overline n}\rangle_\epsilon$ are introduced through the position
\bea
|{\overline n}\rangle_\epsilon &=& (a^\dagger)^n |0\rangle_\epsilon.
\eea
In the above formula, due to the previous remark on the redundancy of the $osp(2|2)$ superalgebra, the creation
operator $a^\dagger$ can denote either $a_1^\dagger$ or $a_2^\dagger$. We denote with $a$ its corresponding annihilation operator satisfying (\ref{kleincomm}).  By exploiting the Klein-deformed commutator (\ref{kleincomm})
and taking into account that $a|0\rangle_\epsilon=0$, we easily obtain the formulas
\bea\label{aziteration}
a|{\overline n}\rangle_\epsilon &=& Z_n |{\overline {n-1}}\rangle_\epsilon, \quad {\text{with}} \quad Z_{2k}=  2k,\quad Z_{2k+1}=2k+1-2\epsilon\beta, \quad k\in {\mathbb N}_0
\eea
(we set, for consistency, $|{\overline 0}\rangle_\epsilon =|0\rangle_\epsilon$).
\par
Let us introduce the normalization coefficients $M_{n,\epsilon}$ through the position
\bea
M_{n,\epsilon}&=& ~_\epsilon\langle{\overline n}|{\overline n}\rangle_\epsilon.
\eea
The equality $~_\epsilon\langle{\overline{ n+1}}|{\overline{ n+1}}\rangle_\epsilon=~_\epsilon\langle{\overline n}|aa^\dagger|{\overline n}\rangle_\epsilon=~_\epsilon\langle{\overline n}|(aa^\dagger-a^\dagger a+a^\dagger a)|{\overline n}\rangle_\epsilon$ implies the following recursive relation for $M_{n,\epsilon}$:
\bea
M_{n+1,\epsilon} &=& (1+2\epsilon\beta (-1)^{n+1})M_{n,\epsilon} + Z_n^2M_{n-1,\epsilon}.
\eea
The first few terms are given by
\bea
M_{0,\epsilon} &=& 1,\nonumber\\
M_{1,\epsilon} &=& (1-2\epsilon\beta),\nonumber\\
M_{2,\epsilon} &=&2(1-2\epsilon\beta),\nonumber\\
M_{3,\epsilon} &=& (1-2\epsilon\beta)(6-4\epsilon\beta).
\eea
It is easily shown that the normalization
of the undeformed oscillator is recovered in the limit $\beta\rightarrow 0$, since $M_{n,\epsilon}\rightarrow n!$.\par
The orthornormal eigenstates, denoted as $|n\rangle_\epsilon$, are given by
\bea
|n\rangle_\epsilon &=& \frac{1}{\sqrt{M_{n,\epsilon}}}|{\overline{n}}\rangle_\epsilon.
\eea

The normalization condition can be defined in closed form in terms of the Pochhammer symbol, introduced through the position
\bea
  (x)_n &= &\frac{\Gamma(x+n)}{\Gamma(x)} = x (x+1) (x+2) \cdots (x+n-1),
  \quad n > 0,
  \nonumber\\
  (x)_0 &= &1.
\eea
The repeated use of (\ref{aziteration}) implies
\begin{align*}
  a_I^{2k} |{2k}\rangle_{\epsilon} 
   &= 2k (2k-1-2\epsilon \beta) (2k-2) (2k-3-2\epsilon \beta) \cdots 
      2 (1-2\epsilon \beta) |{0}\rangle_{\epsilon}=
      \\
   &= (2k)!! \frac{ (-2\epsilon \beta)_{2k} }{ (2k-2-2\epsilon\beta) (2k-4-2\epsilon\beta) (2-2\epsilon\beta) (-2\epsilon \beta)} |{0}\rangle_{\epsilon}=
   \\
   & = \frac{(2k)!!}{2^k} \frac{(-2\epsilon\beta)_{2k}}{(-\epsilon\beta)_k} |{0}\rangle_{\epsilon}= \frac{ k! (-2\epsilon\beta)_{2k}}{(-\epsilon\beta)_k} |{0}\rangle_{\epsilon}.
\end{align*}
It therefore follows that
\bea
  M_{2k,\epsilon} &=& \frac{ k! (-2\epsilon\beta)_{2k}}{(-\epsilon\beta)_k}.
\eea
We furthermore have, also from (\ref{aziteration}),
\bea
  M_{2k+1,\epsilon} &=& (2k+1-2\epsilon\beta) M_{2k,\epsilon} = 
  \frac{k!}{2} \frac{(-2\epsilon\beta)_{2k+2}}{(-\epsilon\beta)_{k+1}}.
\eea

\section{The $n=2$ case with Klein deformed oscillators and $D(2,1;\alpha)$ spectrum-generating superalgebras}

The $n=2$ case corresponds to the $4\times4$ matrix oscillator. With respect to the $n=1$ case, the following features are encountered for the Klein deformation:
\begin{description}
\item[i)] in the presence of a non-vanishing Klein deformation the $osp(4|2)$ spectrum-generating superalgebra of the undeformed case is deformed into a $D(2,1;\alpha)$ spectrum-generating superalgebra; 
\item[ii)] all eigenstates of the model are doubly degenerate; 
\item[iii)] the $D(2,1;\alpha)$ superalgebra is redundant to determine the spectrum of the theory since the ray vectors
of the Hilbert space are determined by a $osp(2|2)$ subalgebra. It provides, nevertheless, a further information due to the fact that $\alpha$ is related with both the Klein deformation parameter and the vacuum energy of the model.
 \end{description}
The operators are explicitely constructed in terms of the $\gamma_J$ ($J=1,2,\ldots,5$) gamma matrices which can be introduced as follows
\bea\label{thegammas}
&\gamma_1 =\sigma_2\otimes\sigma_1 , \quad\gamma_2 =\sigma_2\otimes\sigma_2 ,\quad  \gamma_3 =
\sigma_2\otimes\sigma_3 , \quad\gamma_4 =\sigma_1\otimes{\mathbb I}_2 , \quad\gamma_5=\sigma_3\otimes
{\mathbb I}_2 .&
\eea
The three $\sigma_i$'s are the Pauli matrices introduced in (\ref{pauli}). The block-diagonal matrix $\gamma_5=\gamma_1\gamma_2\gamma_3\gamma_4$ is the fermion parity operator.\par
The eight Hermitian odd operators are $Q_I$, ${\widetilde Q}_I$ ($I=1,2,3,4$). The even Hermitian operators are $H,D,K$, closing the
$sl(2)$ subalgebra, and $S_i, W_{ij}=-W_{ji}$ ($i,j=1,2,3$), closing the $R$-symmetry subalgebra. They are given by
\bea\label{n2spectrumgen}
Q_I&=& \frac{i}{\sqrt{2}}\gamma_I\cdot(\partial_x+\frac{\beta\gamma_5}{x}),\nonumber\\
{\widetilde Q}_I &=& \frac{1}{\sqrt{2}}\gamma_I\cdot x,\nonumber\\
H&=&\frac{1}{2}(-\partial_x^2 +\frac{\beta^2+\beta \gamma_5}{x^2})\cdot{\mathbb I}_4,\nonumber\\ 
D&=& -\frac{i}{2}(x\partial_x+\frac{1}{2})\cdot {\mathbb I}_{4}, \nonumber\\
K&=&\frac{1}{2}x^2\cdot{\mathbb I}_4,\nonumber\\
S_i&=&\frac{i}{2}\gamma_4\gamma_i(1-2\beta\gamma_5),\nonumber\\
W_{ij}&=& \frac{i}{2}\gamma_i\gamma_j(1-2\beta\gamma_5),
\eea
where $\beta$ is the real deformation parameter.\par
Their non-vanishing (anti)commutators realize, see formula (\ref{d21alphaanticomm}), the $D(2,1;\alpha)$ superalgebra with the identification
\bea
\alpha&=&\beta-\frac{1}{2}.
\eea
By repeating the analysis of the $n=1$ case one finds that the ray vectors of the Hilbert space of the model are determined by the $osp(2|2)\subset D(2,1;\alpha)$ subalgebra. Different choices allow to pick up the $osp(2|2)$ spectrum-generating superalgebra. We can, e.g., select the operators to be given by  $H,D,K, Q_1, Q_3, {\widetilde Q}_1, {\widetilde Q}_3, W_{13}$, where the latter operator is the $u(1)$ $R$-symmetry of $osp(2|2)$. An alternative choice consists of the set of operators $H,D,K, Q_2, Q_4, {\widetilde Q}_2, {\widetilde Q}_4 , S_2$.\par
The four pairs of creation/annihilation operators are introduced, as usual, through the positions
$a_{I}= Q_I+i{\widetilde Q}_I$,
$a_{I}^\dagger= Q_I-i{\widetilde Q}_I$. The Klein-deformed commutators now read
\bea
\relax [a_I, a_I^\dagger] &=& {\mathbb I}_4 -2\beta\gamma_5,
\eea
while the $\beta$-deformed oscillator $H_{osc}(\beta)$ is given by
\bea\label{osckleinn2}
&H_{osc}(\beta) = \frac{1}{2}\{a_I, a_I^\dagger\}= H+K=\frac{1}{2}(-\partial_x^2 +x^2+\frac{\beta^2+\beta \gamma_5}{x^2})\cdot{\mathbb I}_4.&
\eea
The commutators
\bea
\relax [H_{osc}(\beta), a_I] =-a_I, &\quad& [H_{osc}(\beta),a_I^\dagger]=a_I^\dagger
\eea
are satisfied.\\
Four lowest weight vectors $|lwv\rangle$ are determined by the condition $a_I|lwv\rangle =0$ for  $I=1,2,3,4$.

The creation operators $a_I^\dagger$ close the ``soft" supersymmetry algebra (see \cite{cuhoto})
\bea
\{a_I^\dagger,a_J^\dagger\}= \delta_{IJ} Z,\quad\quad I,J=1,2,3,4, &\quad&
\relax [Z, a_I^\dagger] = 0,
\eea
where
\bea
Z&=& 2H-2K+4iD
\eea
is a ladder operator.\par
The special points $\alpha=0,-1$ ($\beta=\pm\frac{1}{2}$) correspond (see the comment in Appendix {\bf B})  to the
spectrum-generating superalgebra
\bea
&A(1,1)\oplus su(2)\quad\quad\quad\quad (\text{at}\quad \alpha=0,-1).&
\eea
The selection of the Hilbert space follows the construction for the $n=1$ case.  We have that
\begin{description}
\item[i)] in the range  $\beta \leq-\frac{1}{2}$ the admissible Hilbert space corresponds to a direct sum of the two bosonic lowest weight representations. For $\beta<-\frac{1}{2}$ this construction applies to the $D(2,1;\alpha)$ superalgebras with $\alpha$ belonging to the fundamental domains $FD_1$ and $FD_2$ given in (\ref{fd});
\item[ii)] in the range  $\beta \geq\frac{1}{2}$ the admissible Hilbert space corresponds to a direct sum of the two fermionic lowest weight representations. For $\beta>\frac{1}{2}$ this construction applies to $\alpha$ belonging to the fundamental domains $FD_5$ and $FD_6$ given in (\ref{fd});
\item[iii)] in the intermediate range $-\frac{1}{2}<\beta<\frac{1}{2}$ one can select the Hilbert space as given by the direct sum of the four (two bosonic and two fermionic) lowest weight representations. This case applies to $\alpha$ belonging to the fundamental domains $FD_3$ and $FD_4$ of formula (\ref{fd}).
 \end{description}
We now focus on the third case. The four normalized lowest weight vectors $|0\rangle_{\epsilon,\rho}$, $\epsilon,\rho=\pm 1$,  are
\bea
|0\rangle_{1,1}=N_1 {\small\left(\begin{array}{c} x^{-\beta} e^{-\frac{1}{2}x^2}\\0\\0\\0\end{array}\right)}, &\quad&
|0\rangle_{1,-1}=N_1 {\small\left(\begin{array}{c} 0\\x^{-\beta} e^{-\frac{1}{2}x^2}\\0\\0\end{array}\right)},
\nonumber\\
|0\rangle_{-1,1}=N_{-1} {\small\left(\begin{array}{c}0\\0\\ x^{\beta} e^{-\frac{1}{2}x^2}\\0\end{array}\right)}, &\quad&
|0\rangle_{-1,-1}=N_{-1} {\small\left(\begin{array}{c} 0\\0\\0\\x^{\beta} e^{-\frac{1}{2}x^2}\end{array}\right)}.
\eea
The states $|0\rangle_{1,\rho}$ ($|0\rangle_{-1,\rho}$) are bosonic (fermionic). The normalization factors $N_{\pm 1}$ have been introduced in (\ref{npm1}). The degeneracy of the bosonic (fermionic) energy eigenstates is removed by the eigenvalues of, let's say, the $S_2$ operator ($[S_2,\gamma_5]=[S_2,H_{osc}(\beta)]=0$).\par
By taking into account the $D(2,1;\alpha)$ redundancy, the Hilbert space is spanned by the following ray vectors
which correspond to energy eigenstates
\bea
(a_1^\dagger)^n(a_3^\dagger)^m |0\rangle_{\epsilon,\rho} &=& |{\overline{n,m}}\rangle_{\epsilon,\rho},\quad\quad
n,m\in {\mathbb N}_0.
\eea
Their corresponding energy eigenvalues are
\bea 
E_{n,m;\epsilon,\rho} &=&\frac{1}{2}-\epsilon\beta+ n+m.
\eea
The orthonormalized eigenvectors $|n,m\rangle_{\epsilon,\rho}$ are determined by applying the same techniques as in the $n=1$ case.\par
A similarity transformation, analogous to (\ref{simtra}), is induced by the operator $\gamma_4$. Let $g$ denote an operator of (\ref{n2spectrumgen}). The similarity transformation is defined by
\bea
g&\mapsto& g'=\gamma_4 g\gamma_4.
\eea
In particular
\bea
H'_{osc}(\beta) &=& H_{osc}(-\beta).
\eea
Without loss of generality we can restrict $\beta$ to the non-negative axis $\beta\geq 0$. For the third choice of the Hamiltonian the range $0<\beta<\frac{1}{2}$ corresponds to the (\ref{fd}) fundamental domain $FD_4$ ($-\frac{1}{2}<\alpha<0$) for $\alpha$. In this interval the lowest weight vectors $|0\rangle_{1,\pm 1}$ are the two degenerate bosonic vacua of the theory. The corresponding vacuum energy, expressed in terms of $\alpha$, is
\bea
E_{vac}&=& -\alpha.
\eea
Even if $D(2,1;\alpha)$ is redundant as a spectrum-generating superalgebra, it encodes an important dynamical information of the theory. \par
We point out, as a final remark, that since $\alpha$ belongs to a fundamental domain, all inequivalent (for $\alpha$ real) $D(2,1;\alpha)$ superalgebras are spectrum-generating superalgebras of an associated dynamical system. Stating otherwise, there is no gap in the $\alpha$-induced spectrum generating superalgebras.

\section{The $n=2$ case with non-Klein deformed oscillators and $osp(2|2)$
spectrum-generating superalgebra}

The next construction presents a non-Klein deformation  of the $4\times 4$ matrix oscillator. For this deformation the $osp(4|2)$ spectrum generating superalgebra of the undeformed case  is broken to a $osp(2|2)$ spectrum-generating superalgebra.\par
In this construction the block-antidiagonal, Hermitian, constant matrices $M_I$ entering the (\ref{defsusyop}) deformed supersymmetry operators are different from the ones expressed by (\ref{kleindefsol}). The $M_I$'s are given by a linear
combination $M_I = \nu{\widetilde \gamma}_I+ib\gamma_I\gamma_5$,  where ${\widetilde \gamma}_I$ denotes, up to a sign, one
of the gamma matrices (different from $\gamma_I$ and $\gamma_5$) entering (\ref{thegammas}).  The requirement that the constraints
(\ref{Mrelations},\ref{Vresult},\ref{Dconstraint}) have to be satisfied implies that at most two deformed supersymmetry operators can be constructed,
so that $I=1,2$. The requirement that $V_I=\frac{1}{2}(M_I^2-i\gamma_IM_I)$ is a diagonal matrix and, furthermore,
$V_1=V_2$,  implies that $b$ has to be set to the value $b=\frac{1}{2}$, while $\nu$ is an arbitrary real number.\par
It is easily proven that, without loss of generality (the other solutions being recovered from similarity transformations), an explicit expression of  $Q_1, Q_2$ is given by
\bea
Q_1= \frac{i}{\sqrt 2} (\gamma_1\partial_x -i\frac{M_1}{x}), &&{\text{with}}\quad\quad M_1=\nu\gamma_2+\frac{i}{2}\gamma_1\gamma_5,\nonumber\\
Q_2= \frac{i}{\sqrt 2} (\gamma_3\partial_x -i\frac{M_2}{x}), &&{\text{with}}\quad\quad M_2=-\nu\gamma_4+\frac{i}{2}\gamma_3\gamma_5.
\eea
The $\gamma_I$ matrices were introduced in (\ref{thegammas}). \\
Besides $Q_1,Q_2$, the remaining operators entering the $osp(2|2)$ superalgebra are  $H,K,D, J, {\widetilde Q}_1, {\widetilde Q}_2$. We have, in particular,
\bea
H &=& -\frac{1}{2}\partial_x^2\cdot{\mathbb I}_4+\frac{1}{x^2} V,
\eea
where
\bea
V&=&\frac{1}{8}diag(4\nu^2+8\nu+3,4\nu^2-8\nu+3,4\nu^2-1,4\nu^2-1)
\eea
and
\bea
J&=& -ie_{34}+i e_{43}
\eea
($e_{ij}$ denotes the matrix with entry $1$ at the $i$-th row and $j$-th column and $0$ otherwise). \par 
The operators
$D,K, {\widetilde Q}_1, {\widetilde Q}_2$ are unaffected by the $M_I$'s deformations. Within our conventions we have
\bea
&D= -\frac{i}{2}(x\partial_x+\frac{1}{2})\cdot {\mathbb I}_4,\quad K=\frac{1}{2}x^2\cdot{\mathbb I}_4, \quad
{\widetilde Q}_1 = \frac{1}{\sqrt 2} x \gamma_1, \quad {\widetilde Q}_2 =\frac{1}{\sqrt 2} x \gamma_3.&
\eea
 The non-vanishing (anti)commutators coincide (for the new identification of the corresponding $osp(2|2)$ operators)
with the ones given in (\ref{osp22structureconstants}).\par
Two $a_I, a_I^\dagger$ ($I=1,2)$ pairs  of deformed Heisenberg oscillators are introduced through
\bea
a_I = Q_I+i{\widetilde Q}_I, && a_I^\dagger = Q_I-i{\widetilde Q}_I.
\eea
They define the deformed Heisenberg algebras
\bea
\relax [a_I,a_I^\dagger]&=& {\mathbb I}_4 +G_I,
\eea
where
\bea G_1=diag(-1-2\nu,-1+2\nu,1-2\nu,1+2\nu),&&
 G_2=diag(-1-2\nu,-1+2\nu,1+2\nu,1-2\nu).\nonumber\\&&
\eea
Since $G_I^2$ is not proportional to ${\mathbb I}_4$, these Heisenberg deformations are not of Klein type. One should also note that $G_1\neq G_2$.\par
The deformed oscillator $H_{osc}$ is
\bea\label{nklosc}
&H_{osc}=H+K=\frac{1}{2}\{a_1,a_1^\dagger\}=\frac{1}{2}\{a_2,a_2^\dagger\}=\frac{1}{2}(-\partial_x^2+x^2)\cdot{\mathbb I}_4+\frac{1}{x^2} V .&
\eea
It follows, by taking into account the vanishing anticommutators
\bea
&\{ G_I, a_I\}=\{G_I, a_I^\dagger\}=0,&
\eea
that $a_I$ ($a_I^\dagger$) are annihilation (creation) operators satisfying
\bea
\relax [ H_{osc}, a_I]=-a_I, && [H_{osc},a_I^\dagger]=a_I^\dagger.
\eea
The bosonic (fermionic) states are the eigenfunctions of $\gamma_5$ with eigenvalue $+1$ ($-1$).\par
We are now in the position to introduce the lowest weight representations. A lowest weight vector $|lwv\rangle$ satisfies the condition
\bea
a_I|lwv\rangle &=&0, \quad {\text{for}}\quad I=1,2.
\eea
Two (both bosonic) lowest weight vectors are found. They are given by $\Psi_{1,2}(x)$, where
\bea
\Psi_1(x)={\small\left(\begin{array}{c} x^{-(\frac{1}{2}+\nu)} e^{-\frac{1}{2}x^2}\\0\\0\\0\end{array}\right)}, &\quad&
\Psi_2(x)={\small\left(\begin{array}{c} 0\\x^{(\nu-\frac{1}{2})} e^{-\frac{1}{2}x^2}\\0\\0\end{array}\right)}.
\eea
The fermionic states $\Psi_3(x)= a_1^\dagger \Psi_1(x)$, $\Psi_4(x)=a_1^\dagger \Psi_2(x)$, $\Psi_5(x)=a_2^\dagger
\Psi_1(x)$, $\Psi_6(x)=a_2^\dagger \Psi_2(x)$ satisfy the 
conditions $a_2\Psi_3(x)=a_2\Psi_4(x)=a_1\Psi_5(x)=a_1\Psi_6(x)=0$. They are, nevertheless, excited states belonging
to the lowest weight representations induced by $\Psi_{1,2}(x)$.\par
As recalled in Appendix {\bf C}, a wavefunction of the form $x^\beta e^{-\frac{1}{2}x^2}$ is normalized provided that
$\beta>-\frac{1}{2}$. It follows that a normalized lowest weight vector is encountered, provided that
\bea
\nu \neq 0.
\eea
In this range the Hilbert space is given by a single lowest weight representation. For $\nu<0$ the normalizable lowest weight state is $\Psi_1(x)$; for $\nu>0$ the normalizable lowest weight state is $\Psi_2(x)$.  The vacuum energy $E_{vac}$,
in the admissible $\nu\neq 0$ range, is
\bea
E_{vac} &=& -\frac{1}{2}+|\nu|.
\eea
The spectrum of the theory is given by
\bea
E_n &=& -\frac{1}{2}+|\nu|+n,\quad\quad n\in {\mathbb N}_0.
\eea
With the exception of the single vacuum state, all excited states for $n\geq 1$ are doubly degenerate. 
This follows from the ${\cal N}=2$ soft supersymmetry algebra satisfied by the two creation operators $a_I^\dagger$,
given by
\bea\label{softsusyn2}
\{a_I^\dagger,a_J^\dagger\}= \delta_{IJ} Z,\quad\quad  &\quad&
\relax [Z, a_I^\dagger] = 0,
\eea
with
\bea
Z&=& 2H-2K+4iD.
\eea
Therefore, the spectrum corresponds to the semi-infinite tower of $(1,2,2,2,\ldots )$ states.\par
Let us consider the $\nu>0$ case. In this case the normalized vacuum state $|0\rangle$ is
\bea
|0\rangle&=& N\Psi_2(x), \quad\quad{\text{with}}\quad \quad N=\frac{1}{{\sqrt{\Gamma (\nu)}}}.
\eea
The $n$ excited states are spanned by the vectors $(a_1^\dagger)^{n_1}(a_2^\dagger)^{n_2}|0\rangle$, where
$n=n_1+n_2$.  At given $n>0$, due to the (\ref{softsusyn2}) relation, only two of the associated ray vectors are distinct. They can be chosen to be expressed through
\bea
|{\overline {n,0}}\rangle =  (a_1^\dagger)^{n}|0\rangle, &\quad& 
|{\overline {n-1,1}}\rangle =  (a_1^\dagger)^{n-1}a_2^\dagger|0\rangle.
\eea
By applying the method discussed in Section {\bf 3} we can compute the orthonormal states for $\nu>0$.
We report here just the final results. The orthonormal states are
\bea
|n,0\rangle = N_{n,0}|{\overline {n,0}}\rangle, &\quad& 
|n-1,1\rangle = N_{n-1,1}|{\overline {n-1,1}}\rangle,
\eea
where
\begin{align}
   N_{n,0} &= \left( 2^n \left\lfloor \frac{n}{2} \right\rfloor ! \, (\nu)_{\lceil \frac{n}{2} \rceil} \right)^{-1/2},\nonumber
   \\
   N_{2m-1,1} &= \frac{1}{ \sqrt{2^{2m} (m-1)!\, (\nu)_{m+1}}},
    \qquad
    N_{2m,1} = \frac{1}{\sqrt{2^{2m+1} m!\, (\nu)_{m+1}}}.
\end{align}
In the above equations $ (\nu)_m $ denotes the Pochhammer symbol, while $ \lfloor x \rfloor $ and $ \lceil x \rceil $ are, respectively,  the floor and ceiling functions.

\section{Comments on the general $n$ case}

In Section {\bf 3} we presented the conditions to be satisfied in order to have a spectrum-generating superalgebra
for the inverse-square potential deformed matrix oscillators. A scale-invariant supersymmetric quantum mechanics is implied by
fulfilling the conditions (\ref{Mrelations}) and (\ref{Vresult}). The existence of a spectrum-generating superconformal algebra is further implied by satisfying (\ref{Dconstraint}), plus the requirement for the fermionic generators to belong to a representation multiplet of the $R$-symmetry generators (\ref{sigmadef}). We presented the most general solutions (up to similarity transformations) for $n=1$ (in Section {\bf 4}) and
$n=2$ (in Sections {\bf 5} and {\bf 6} for deformations of, respectively, Klein type and non-Klein type). \par
It is beyond the scope of the present paper to investigate the most general class of solutions for $n\geq 3$. This will be left to future works. It is worth, nevertheless, to introduce the present state of the art and discuss some general
features which can be noted.  Beyond $n=2$, a non-trivial solution was found in \cite{akt} for $n=4$ (the associated spectrum-generating superalgebra being $F(4)$, with ${\cal N}=8$ supersymmetries, see Appendix {\bf A}). \par
So far this is the only known  non-trivial solution for $n\geq 3$. Its construction was made possible by the large
symmetry of the model, reflected by the so-called ``octonionic covariance" which, essentially, derives from the construction of its gamma matrices in terms of the octonionic structure constants. The model is unique (up to similarity transformations) and corresponds, even if not explicitly stated in \cite{akt}, to a deformation of non-Klein type.  The results of \cite{akt} rule out $n=4$ non-trivial octonionic covariant deformations based on the ${\cal N}=8$ superconformal algebra $osp(8|2)$ and on the ${\cal N}=7$ exceptional superconformal algebra $G(3)$.\par
The following picture emerges:
\begin{description}
\item[i)] at $n=1$ the deformation is of Klein type and depends on a real continuous parameter $\beta$. In the 
$\beta\rightarrow 0$ limit the undeformed oscillator is recovered. The spectrum-generating superalgebra $osp(2|2)$
is recovered for both undeformed and deformed oscillators;
\item[ii)] for $n=2$ two new features appear. The deformation of Klein type, which depends on a continuous parameter $\alpha$, is such that its spectrum-generating superalgebra is deformed, since the $osp(4|2)$ spectrum-generating superalgebra of the undeformed oscillator (recovered in the $\alpha\rightarrow 0$ limit) is replaced by $D(2,1;\alpha)$. The second new feature is the appearance of the non-Klein deformation which depends on a continuous
parameter $\nu\neq 0$. Contrary to the Klein type deformation, the non-Klein deformation is not connected with
the undeformed oscillator;
\item[iii)] at $n=4$ the non-Klein deformation possesses the spectrum-generating superalgebra $F(4)$ and is point-like. It corresponds to an isolated point of the inverse-square potential coupling constants entering the diagonal matrix Hamiltonian. The deformation is obviously not
connected with the $osp(8|2)$ undeformed oscillator.
 \end{description}

\section{Conclusions}

The systematic construction of inverse-square potential deformed matrix oscillators with superconformal
spectrum-generating superalgebras  for larger ($n\geq 3$) matrices is left for future works (the only case which is known, the \cite{akt}
construction for $n=4$ and $F(4)$ superalgebra, is made possible by the simplifications due to its huge symmetry).
In a forthcoming work we will present the results for $d$-dimensional, with $d\geq 2$, deformed matrix oscillators. 
Another promising future line of research consists in addressing the multi-particle case. It requires extending at the quantum level the construction which is done (see, eg., the recent \cite{feiv} paper) for multi-particle classical superconformal world-line models. \par
We conclude with two more comments. The first one is the recognition that, since in a certain range of the deformation parameter
the  Hilbert space can be taken as a direct sum of lowest weight representations of its spectrum-generating superalgebra, therefore the superalgebra does not contain all information about the spectrum of the theory (not every higher energy excited state
is connected to a given lower energy state via superalgebra ladder operators). This offers the tantalizing possibility that extra algebraic structures, possibly infinite-dimensional, could be responsible for that and used to generate the whole spectrum  of the theory.\par
The final comment concerns the possible interesting applications of  these models to higher-spin theories (as recognized in \cite{tv}), in a implementation of the AdS/CFT holography. This is based on the property that Klein-deformed oscillators with $osp(2|2)$ spectrum-generating superalgebra provide a realization of the Vasiliev's higher spin superalgebra introduced in \cite{vas}. Recently,
 the relevance of non-Klein deformed oscillators to higher spin theories was pointed out (see e.g. \cite{vas2} and references therein).

~
\par
~\\
{\bf{\Large{Appendix A: the $1D$ finite superconformal Lie algebras}}}
~\par
~\par
The set of the one-dimensional finite superconformal Lie algebras is a subclass of the finite simple Lie superalgebras
entering the Kac's classification \cite{kac} (see also the classification in \cite{snr1} and, for exceptional superalgebras, \cite{snr2}, as well as the \cite{dic} review) and satisfying the following additional properties \cite{tian}.  Any such Lie superalgebra ${\cal G}$ over the field ${\mathbb C}$ admits a $5$-grading decomposition
\bea
{\cal G} &=& {\cal G}_{-1}\oplus {\cal G}_{-\frac{1}{2}}\oplus
{\cal G}_0\oplus {\cal G}_{\frac{1}{2}}\oplus {\cal G}_{1}.
\eea
The (anti)commutators (compactly denoted as ``$\relax [. ,. \}$") satisfy the condition
\bea
[{\cal G}_i,{\cal G}_j\}& \subset& {\cal G}_{i+j}.
\eea
The even sector ${\cal G}_{even}={\cal G}_{-1}\oplus{\cal G}_0\oplus{\cal G}_1$ is isomorphic to the direct sum of the Lie algebras $sl(2)\oplus R$, where the subalgebra $R$ is known as $R$-symmetry. \\
The odd sector ${\cal G}_{odd}={\cal G}_{-\frac{1}{2}}\oplus {\cal G}_{\frac{1}{2}}$ is spanned by $2{\cal N}$ generators. Accordingly, each finite superconformal Lie algebra ${\cal G}$ is labeled  by its associated positive integer ${\cal N}$.\\
The positive sector ${\cal G}_{>0}$ is isomorphic to the algebra of the ${\cal N}$-extended supersymmetric quantum mechanics \cite{{wit},{dcr}} defined by the (anti)commutators
\bea\label{sqm}
&\{Q_I, Q_J\}= 2\delta_{IJ} H, \quad\quad [H,Q_I] =0,\quad\quad  {\text{for}}\quad I,J=1,\ldots, {\cal N}.&
\eea
The generator $H$ is the positive root of the $sl(2)$ subalgebra. The $sl(2)$ Cartan and negative root generators are denoted as $D$, $K$, respectively. The negative sector ${\cal G}_{<0}$ satisfies the subalgebra
\bea\label{confsqm}
&\{{\widetilde Q}_I,{\widetilde Q}_J\}= 2\delta_{IJ} K, \quad\quad [K,{\widetilde Q}_I] =0,\quad\quad  {\text{for}}\quad I,J=1,\ldots, {\cal N}.&
\eea
The sector ${\cal G}_1$ (${\cal G}_{-1}$) is spanned by $H$ ($K$); the odd sector
${\cal G}_{\frac{1}{2}}$ (${\cal G}_{-\frac{1}{2}}$) is spanned by the $Q_I$ (${\widetilde Q}_I$) generators; finally, the ${\cal G}_0$ sector is ${\cal G}_0=D{\Bbb C}\oplus R$.\par
In this paper  on spectrum-generating superalgebras of the matrix inverse-square potential models several examples of one-dimensional conformal Lie superalgebras appear. In particular we mentioned superalgebras belonging to the classical series, such as $D({\cal N},1)\sim osp(2{\cal N}|2)$ (defined for any positive ${\cal N}$ and with
$so(2{\cal N})$ as associated R-symmetry) and $B(n,1)\sim osp(2n+1|2)$ (such that
${\cal N}=2n+1$ and with $so(2n+1)$ as R-symmetry), as well as the exceptional superalgebras $D(2,1;\alpha)$ (superconformal for ${\cal N}=4$ and discussed in Appendix {\bf B}), $G(3)$ and $F(4)$. The latter two exceptional superalgebras are superconformal for, respectively, ${\cal N}=7$ with $g_2$ as R-symmetry and ${\cal N}=8$ with $so(7)$ as R-symmetry.  \par
The list of one-dimensional superconformal Lie algebras further includes $A(n,1)$, $D(2,n)$ (see \cite{dic} for their definition). The complete list of one-dimensional superconformal Lie algebras with ${\cal N}\leq 8$ is presented in \cite{tian}.\par
We present here for convenience the non-vanishing (anti)commutators of the $osp(2|2)$ superalgebra generators given by the (\ref{osp22op}) operators, with $I,J=1,2$. We have
\bea\label{osp22structureconstants}
\relax& [D,K]=-iK,\quad\quad\quad\quad\quad [D,H]=iH,\quad\quad\quad\quad\quad \quad~~[H,K]=-2i D,\quad\quad\quad\quad&\nonumber\\
\relax&~~~~~ \{Q_I,Q_J\} = 2\delta_{IJ} H,\quad\quad\quad ~\{{\widetilde Q}_I, {\widetilde Q}_J\}= 2\delta_{IJ}K,\quad\quad\quad\quad
\{Q_I, {\widetilde Q}_J\} = -2\delta_{IJ}D +\epsilon_{IJ}J,~~~~&\nonumber\\
\relax &[D, Q_I]=\frac{i}{2}Q_I,\quad\quad \quad\quad\quad\quad\quad[D, {\widetilde Q}_I]= -\frac{i}{2}{\widetilde Q}_I,\quad\quad \quad\quad\quad\quad&\nonumber\\
\relax &[Q_I, K] = i{\widetilde Q}_I,\quad\quad\quad \quad\quad\quad\quad [{\widetilde Q}_I, H]=-i Q_I,\quad\quad \quad\quad\quad\quad&\nonumber\\
\relax&\quad \quad\quad\quad\quad\quad[J, Q_I] = -i \epsilon_{IJ} Q_J,  \quad\quad\quad\quad\quad[J, {\widetilde Q}_I]= -i \epsilon_{IJ} {\widetilde Q}_J. \quad\quad\quad\quad\quad\quad\quad\quad\quad\quad&
\eea
In the above relations $\epsilon_{12}=-\epsilon_{21}=1$ is the totally antisymmetric tensor.\par
~
\par
~\\
{\bf{\Large{Appendix B: basic properties of the $D(2,1;\alpha)$ superalgebras}}}
~\par
~\par
The exceptional superalgebras $D(2,1;\alpha)$ are parametrized, see \cite{dic}, by $\alpha\in{\mathbb C}\setminus  \{0,-1\}$,
with $\alpha$ entering the structure constants. As recalled in Appendix {\bf A}, they are ${\cal N}=4$ superconformal Lie algebras. Their even sector ${\cal G}_{even}$ is the direct sum of three $sl(2)$ subalgebras, so that
\bea
{\cal G}_{even} &=& sl(2)\oplus sl(2)\oplus sl(2).
\eea
The three $sl(2)$ subalgebras can be interchanged. As a result, the $S_3$ permutation group of three elements acts on $\alpha$; two generators of $S_3$ are expressed as the transformations $\alpha\mapsto \frac{1}{\alpha}$, $\alpha\mapsto -(1+\alpha)$.  An $S_3$-orbit is given by the following set of elements
\bea
&\{\alpha,~ ~{\frac{1}{\alpha}},~~-(1+\alpha),~~-\frac{1}{(1+\alpha)},~~-\frac{(1+\alpha)}{\alpha}, ~~-\frac{\alpha}{(1+\alpha)}\} .&
\eea
The superalgebras specified by $\alpha, \alpha'$ belonging to the same $S_3$-orbit are isomorphic. \par
The special values 
\bea 
\alpha &=&-2,~-\frac{1}{2},~ ~1
\eea 
correspond to the superalgebra $D(2,1)\sim osp(4|2)$ which belongs to the classical series of orthosymplectic superalgebras.
\par
The structure constants can also be defined at the special values $\alpha=0,-1$. For these values, on the other hand, the superalgebra is no longer simple, being given by the direct sum $A(1,1)\oplus sl(2)$ (the generators of one of the three $sl(2)$ subalgebras decouple from the remaining generators). The simple superalgebra $A(1,1)$ is ${\cal N}=4$ superconformal. \par
The hermiticity property of the Hamiltonians of the matrix inverse-square potential quantum mechanics (both in presence or in absence of the oscillatorial term) requires $\alpha$ to be real. For $\alpha\in{\mathbb R}$, the following six fundamental domains under the group of $S_3$ transformations are encountered \cite{kuto}:
\bea&&\label{fd}
\begin{array}{ccrcccr}\\
FD_1:&\quad&-\infty&<&\alpha&\leq& -2,\\
FD_2:&\quad&-2&\leq&\alpha& < &-1,\\
FD_3:&\quad&-1&<&\alpha &\leq &-\frac{1}{2},\\
FD_4:&\quad&-\frac{1}{2}&\leq&\alpha &< &0,\\
FD_5:&\quad&0 &<&\alpha &\leq &1,\\
FD_6:&\quad&1&\leq&\alpha& <& \infty.
\end{array}
\eea
~\par
The operators given in formula (\ref{n2spectrumgen}) produce the $D(2,1;\alpha)$ superalgebra; their non-vanishing (anti)commutators are given by
\bea\label{d21alphaanticomm}
\relax[D,K]&=&-iK,\nonumber\\
\relax [D,H]&=&iH,\nonumber\\
\relax [H,K]&=&-2i D,\nonumber\\
 \{Q_I,Q_J\} &=& 2\delta_{IJ} H,\nonumber\\
\relax\{{\widetilde Q}_I, {\widetilde Q}_I\}&=& 2\delta_{IJ}K,\nonumber\\
\relax\{Q_I,{\widetilde Q}_I\}&=& -2D,\nonumber\\
\{Q_4, {\widetilde Q}_i\} &=&S_i,\nonumber\\
\{{\widetilde Q}_4,Q_i\} &=&- S_i,\nonumber\\
\{Q_i,{\widetilde Q}_j\}&=& W_{ij},\nonumber\\
\relax [D, Q_I]&=&\frac{i}{2}Q_I,\nonumber\\
\relax [D, {\widetilde Q}_I]&=& -\frac{i}{2}{\widetilde Q}_I,\nonumber\\
\relax [Q_I, K] &=& i{\widetilde Q}_I,\nonumber\\
 \relax [{\widetilde Q}_I, H]&=&-i Q_I,\nonumber\\
\relax[Q_4, S_i] &=& i Q_i,\nonumber\\
\relax [{\widetilde Q}_i, S_j]&=& -i \delta_{ij} {\widetilde Q}_4 + 2i\beta \epsilon_{ijk} {\widetilde Q}_k,\nonumber\\
\relax[Q_i,S_j]&=& -i \delta_{ij} Q_4 + 2i\beta \epsilon_{ijk} Q_k,\nonumber\\
\relax[{\widetilde Q}_4, W_{ij} ] &=& -2i \beta \epsilon_{ijk} {\widetilde Q}_k,\nonumber\\
\relax[Q_4, W_{ij}] &=& -2i \beta \epsilon_{ijk} Q_k, \nonumber\\
\relax[{\widetilde Q}_i, W_{jk}]&=& i ( \delta_{ij} {\widetilde Q}_k - \delta_{ik} {\widetilde Q}_j) + 2i \beta \epsilon_{ijk} {\widetilde Q}_4,\nonumber\\
\relax[Q_i, W_{jk}]&=& i ( \delta_{ij} Q_k - \delta_{ik} Q_j) + 2i \beta \epsilon_{ijk} Q_4,\nonumber\\
\relax[S_i,S_j]&=& -i W_{ij} + 2 i \beta \epsilon_{ijk} S_k,\nonumber\\
\relax[S_i, W_{jk}] &=& i \delta_{ij} (S_k - \beta \epsilon_{k\ell m} W_{\ell m}) - i \delta_{ik} (S_j - \beta \epsilon_{j\ell m} W_{\ell m}),\nonumber\\
\relax[W_{ij},W_{kl}]&=& i ( \delta_{ik} \widetilde{W}_{\ell j} - \delta_{i \ell} \widetilde{W}_{kj} + \delta_{jk} \widetilde{W}_{i \ell} - \delta_{j\ell} \widetilde{W}_{ik}),
\eea
where $ \widetilde{W}_{ij} = W_{ij} - 2 \beta \epsilon_{ijk} S_k$.  
The deformation parameter $\beta$ entering the (\ref{n2spectrumgen}) operators and the above (anti)commutators
is related to $\alpha$ through the equation
$\alpha=\beta-\frac{1}{2}$.
~\par

\par
~\\
{\bf{\Large{Appendix C: selecting the Hilbert spaces of the models}}}
~\par
~\par
The selection of the viable Hilbert spaces of the matrix superconformal quantum mechanics (with or without the addition of the extra oscillatorial term) requires the preliminary knowledge of the selection of the Hilbert spaces for
either the one-dimensional inverse-square potential model \cite{cal} defined by the Hamiltonian
\bea\label{hcal}
H_{def} &=& \frac{1}{2}\left(-\partial_x^2+\frac{g}{x^2}\right)
\eea
or the de Alfaro-Fubini-Furlan model \cite{dff} defined by the Hamiltonian
\bea\label{hdff}
H_{DFF} &=& \frac{1}{2}\left(-\partial_x^2+\frac{g}{x^2}+x^2\right).
\eea
An extensive analysis of the admissible choices of their Hilbert spaces for $g>0$ was given in \cite{{mt,ftf}}. We present here, in a slightly modified form suitable for our purposes, the results of \cite{{dff,mt,ftf}} concerning the choice of the Hilbert spaces for the $H_{DFF}$ (\ref{hdff}) Hamiltonian.\par
Following \cite{op}, the ground state wave function of $H_{DFF}$ has the form
\bea
\Psi_\beta &=& x^\beta e^{-\frac{1}{2}x^2}.
\eea
$\Psi_\beta(x)$ is an eigenfunction (not necessarily the groundstate) of $H_{DFF}$ provided that the relation
\bea\label{betag}
g&=& \beta^2-\beta
\eea
holds. Its associated energy eigenvalue  $E_\beta$ is 
\bea
E_\beta &=&\frac{1}{2}+\beta.
\eea
The two solutions of the (\ref{betag}) equation are $\beta_\pm$, given by
\bea
\beta_\pm &=& \frac{1\pm\sqrt{1+4g}}{2}.
\eea
The reality of $E_\beta$ requires $\beta$ to be real; therefore the coupling constant $g$ needs to be
\bea g&\geq& -\frac{1}{4}.
\eea
The wave function $\Psi_\beta(x)$ is square integrable in the real line provided that
\bea\label{normalizedcondition}
&\int_{-\infty}^{+\infty}dx|\Psi_\beta (x)|^2 =\int_{-\infty}^{+\infty}dx x^{2\beta}e^{-{x^2}}=C_\beta <\infty.  &
\eea
Taking into account the singularity at the origin for negative $\beta$, the  above condition is satisfied for
\bea
\beta&> & -\frac{1}{2}.
\eea

In the range
\bea
\beta&>&0
\eea
the wavefunction $\Psi_\beta(x)$ can be defined in the $x\geq 0$ non negative half line ${\mathbb R}^+$; it satisfies the Dirichlet boundary condition at the origin ($\Psi_\beta(0)=0$). \par
In the range
\bea
-\frac{1}{2}< &\beta &\leq 0
\eea
the wavefunction $\Psi_\beta (x)$ is necessarily defined as ${\cal L}^2({\mathbb R})$ square-integrable function on the real line.\par
The dynamical symmetry of the time-dependent Schr\"odinger equation with (\ref{hcal}) or (\ref{hdff}) as Hamiltonian is, see e.g. \cite{tv}, $sl(2)\oplus u(1) $, so that the $sl(2)$ algebra is the spectrum generating algebra. The wavefunctions
$\Psi_{\beta_\pm}(x)$ are the lowest weight vectors of the $sl(2)$ lowest weight representations associated with $H_{DFF}$. 
All excited states obtained by applying the raising ladder operators to $\Psi_{\beta_\pm}(x)$ belong, for ${\beta_\pm} >0$, to the functions on the half line which satisfy the Dirichlet boundary condition and, for $-\frac{1}{2}<\beta_{\pm}\leq 0$, to the ${\cal L}^2({\mathbb R})$ square-integrable functions on the real line.\par
The Hilbert space of the model is either a single lowest weight representation of $sl(2)$ or the direct sum of its two lowest weight representations. \\
In the admissible $g\geq -\frac{1}{4}$ interval of the coupling constant we have:\\
\begin{description}
\item[i)] at $g=-\frac{1}{4}$, $\beta_+=\beta_-=\frac{1}{2}$, so that there is a single lowest weight representation; its wavefunctions are defined on the half line and satisfy the Dirichlet boundary condition;
\item[ii)]
 in the range $-\frac{1}{4}<g<0$, $\beta_\pm$ are both positive. The Hilbert space is the direct sum of the two lowest weight representations. Its wavefunctions are defined on the half line and satisfy the Dirichlet boundary condition. $\Psi_{\beta_-}(x)$ is the ground state; 
\item[iii)] at $g=0$, $H_{DFF}$ is the Hamiltonian of the ordinary one-dimensional oscillator. 
The two lowest weight representations correspond to wavefunctions which are respectively even (odd) under the 
$x \mapsto -x$ parity transformation. The gaussian $\Psi_{\beta_-}(x)$ is the ground state and the lowest state of the even parity eigenfunctions. The first excited state is given by $\Psi_{\beta_+}(x)$, the lowest weight vector of the odd-parity eigenfunctions;
\item[iv)] in the range $0<g<\frac{3}{4}$, $\beta_+$ is positive while $\beta_-$ is negative. Following \cite{mt,ftf}, two choices of Hilbert space can be made. Either the single lowest weight representation with $\Psi_{\beta_+}(x)$ as lowest weight vector (correponding to functions on the half line with Dirichlet boundary condition at the origin), or the direct sum of the two lowest weight representations corresponding to square integrable functions on the real line. In this latter case $\Psi_{\beta_-}(x)$ is the ground state;
\item[v)] for $g\geq \frac{3}{4}$, since $\beta_-\leq-\frac{1}{2}$, the wavefunction $\Psi_{\beta_-}(x)$ does not satisfy (\ref{normalizedcondition}) and is not normalized. The Hilbert space is given by a single $sl(2)$ lowest weight representation with $\Psi_{\beta_+}(x)$ as  lowest weight vector and ground state. 
 \end{description}
In all above cases the Hamiltonian $H_{DFF}$ is well defined as a self-adjoint operator acting on the corresponding Hilbert space. Its eigenvalues are discrete and bounded from below.

\par

~\par
~\par
\par {\Large{\bf Acknowledgments}}
{}~\par{}~\par

N. A. is supported by the grants-in-aid from JSPS (Contract No. 26400209). He thanks UFABC and CBPF for hospitality.  I. E. C. acknowledges a CNPq Ph.D. grant.  Z. K. and F. T. are grateful to the Osaka Prefecture University for hospitality.  The work was supported by CNPq (PQ grant 308095/2017-0).

\end{document}